# Predictive Software Scheduling as an Early-Warning Hint Layer for Optical Engine Thermal Drift in Heterogeneous SoIC Packaging

Chi Fei Chung

*XRM-SSD Engineering, Dollarchip Technology Inc.*

## Abstract

As semiconductor scaling reaches the A16 / 2 nm node, the integration of co-packaged optics (CPO) via TSMC's Compact Universal Photonic Engine (COUPE) architecture introduces critical thermal-optical coupling challenges. Micro-ring resonators embedded in the Photonic Integrated Circuit (PIC) layer are exquisitely sensitive to temperature: a wafer-level center wavelength deviation of merely ±1.7 nm across the 300 mm wafer represents the COUPE platform's manufacturing control limit. We propose XRM-SSD V24, a physics-aware scheduling layer that models inference-load density 20–50 ms before execution and issues early-warning hints to the COUPE bias-control firmware, enabling pre-emptive thermal compensation. Simulation-based validation on a software emulation platform (physical characterization pending TSMC tape-out) over 90,000 inference steps yields a simulator-internal thermal-load correlation of $R^2 = 0.9911$ across a workload density range $\rho v24 \in [0.9, 2.7]$ (a 3× span) and wavelength drift below 0.354 nm – 21% of the ±1.7 nm wafer-level wavelength control budget and 71% of the tighter ±0.5 nm per-channel spectral specification. A full Thermal Resistance Fingerprint characterization further confirms $R_{th} = 0.45$ °C/W, thermal time constant $\tau = 80$ ms, and thermo-optic coefficient 0.0852 nm/°C across five discrete load states (Idle to Peak). Memory stability is reported as zero leakage in the current simulation run; long-duration soak testing to confirm sustained stability remains future work. We establish a formal domain separation between deterministic software scheduling and continuous physical thermal dynamics, ensuring physics-consistent claims suitable for peer review.

## I. INTRODUCTION

The transition to heterogeneous integration at the 2 nm node marks a decisive inflection point for high-performance AI compute hardware. TSMC's System-on-Integrated-Chips (SoIC) technology stacks an Electronic Integrated Circuit (EIC) — carrying the transformer-based inference accelerator — directly atop a Photonic Integrated Circuit (PIC) housing co-packaged optical transceivers. This vertical proximity creates a tight thermal coupling channel absent in conventional 2.5D packaging.

The COUPE architecture places silicon micro-ring resonators within microns of the active EIC power domain. The thermal time constant of the SoIC package (~80 ms, simulation-confirmed herein) is smaller than the burst duration of modern LLM inference workloads (~100–500 ms), meaning transient compute surges reliably produce measurable wavelength excursions at the photonic layer.

Existing approaches rely on closed-loop feedback: an on-chip temperature sensor triggers a microheater bias adjustment after the thermal event has already propagated. XRM-SSD V24 relocates the compensation decision to the software scheduling layer. By predicting the upcoming thermal load from token-level metadata 20–50 ms before execution, V24 provides a *hardware hint* to the COUPE thermal controller, transforming reactive compensation into pre-emptive stabilization.

This paper contributes:

• A cross-layer thermal coupling model linking inference density (ρv24) to PIC temperature differential (ΔT).

• Simulated validation on a 90,000-step dataset demonstrating $R^2 = 0.9911$.

• A six-panel Thermal Resistance Fingerprint Map characterizing $R_{th}$, T, and thermo-optic behavior across five load states.

• A domain separation framework ensuring physics-consistent peer-review claims.

## II. BACKGROUND AND RELATED WORK

### A. TSMC COUPE Architecture

TSMC's Compact Universal Photonic Engine leverages SoIC face-to-face bonding to achieve sub-2 pJ/bit optical I/O energy efficiency [1]. Silicon micro-ring resonators fabricated on the COUPE platform exhibit a wafer-level center wavelength controlled within ±1.7 nm across the 300 mm wafer [6]. Each resonator is equipped with a local microheater consuming ~10–20 mW per channel at steady state

### B. Thermal-Optical Coupling in SoIC

Thermal crosstalk in vertically-stacked SoIC packages scales inversely with EIC-PIC separation distance d [2]. In COUPE's face-to-face configuration (d ~ 5–15 μm), coupling coefficients Gamma are an order of magnitude larger than lateral-integration approaches, making thermal management non-optional at production inference frequencies.

### C. Thermal Resistance Fingerprinting

Thermal resistance $R_{th} = \Delta T / \Delta P$ characterizes the steady-state heat conduction efficiency of a package. Prior work has applied fingerprint methods to CPU/GPU thermals [3]; application to co-packaged photonic layers — where $R_{th}$ directly determines wavelength budget consumption — is novel. V24's fingerprint validation establishes $R_{th}$ = 0.45°C/W with agreement between theoretical prediction and simulated data across five load states.

## III. CROSS-LAYER THERMAL MODEL

### A. Inference Load Density and System Throughput Mapping

The V24 orchestration layer tracks computing intensity by calculating the instantaneous, dimensionless local workload density ratio ρv24(t), which aggregates structural computational complexity across all N(t) concurrent active execution streams:

$$\rho_{\mathrm{v24}}(t) = \sum_{i=1}^{N(t)} \left[\mathrm{Attn}(i) \cdot w(i) \cdot \mathcal{F}(i)\right]$$

where Attn(i) represents the attention weight matrix footprint, ω(i) denotes the active parameter activation rate, and F(i) is the geometric routing coefficient. To prevent notation collisions with downstream performance metrics, the global system execution performance is decoupled from local density and expressed as the System Throughput Coefficient $T_{24}$ (measured in millions of tasks per second, MTPS). The orchestration layer enforces an explicit affine transformation block to map runtime density to physical execution velocity:

$$T_{24} = \alpha \cdot \rho v_{24}(t) + \beta$$

where the simulated calibration constants α = 0.361 MTPS and β = 19.875 bind a workload density domain of ρv24 ∈ in [0.9, 2.7] to a strict, predictable system throughput domain of [20.20, 20.85] MTPS.

### B. Unified Thermal Convolution Model

Transient heat conduction from the Electronic Integrated Circuit (EIC) layer through the system's System-on-Integrated-Chips (SoIC) interface down to the Photonic Integrated Circuit (PIC) substrate follows a first-order thermal network. Rather than an unweighted power integral, the physical state boundary must include the appropriate steady-state gain normalization to satisfy energy conservation. The time-varying temperature delta $\Delta T_{PIC}(t)$ is governed by the unified convolution model:

$$\Delta T_{\mathrm{PIC}}(t) = \int_0^t \left(\frac{R_{\mathrm{th}} \cdot \Gamma(d)}{\tau_{\mathrm{th}}}\right) \cdot \exp\left(-\frac{t-u}{\tau_{\mathrm{th}}}\right) \cdot \Delta P_{\mathrm{EIC}}(u)\, du$$

where $\tau_{th}$ = 80 ms is the physical hardware thermal RC time constant of the reference silicon package, and Γ(d) is the dimensionless spatial coupling factor across a physical substrate displacement 2D distance matrix *d*, and $R_{th}$ represents the localized junction thermal resistance.

### C. Multilayer Thermal Boundary Characterization

The steady-state thermal resistance configuration is modeled as a sequence of series thermal boundaries moving outward from the silicon junction to the ambient environment. The total system boundary behavior is defined by the cumulative series thermal resistance $\sum R_{th,\,i}$:

$$\sum R_{\mathrm{th},i} = \frac{\Delta T}{\Delta P} \quad [^\circ\mathrm{C/W}]$$

Simulated fingerprint verification of the reference co-packaged silicon package establishes a true localized junction-to-substrate thermal resistance of $R_{th}$ = 0.45 °C/W. Package-level step power changes under active peak load over a representative step power ΔP = 82 W. This step-function drives a linear junction temperature delta:

$$\Delta T = 0.45\ ^\circ\mathrm{C/W} \times 82\ \mathrm{W} = 36.90^\circ\mathrm{C} \approx 37^\circ\mathrm{C}$$

This linear dependency holds within the simulated load states when mapped against the Watts-scaled package power domain [0, 100 W], eliminating visual slope mismatches and anchoring the structural annotations demonstrated in Figure 2. The series boundaries evaluate to distinct cumulative milestones: Junction-to-Case (0.812°C/W), Case-to-Heatsink (1.407°C/W cumulative), and Heatsink-to-Ambient (1.995°C/W total cumulative system boundary). Note that this cumulative junction-to-ambient figure (1.995 °C/W) and the unified $R_{th}$ = 0.45 °C/W validated throughout this paper refer to different reference planes: the former is the full package-to-ambient boundary stack, while the latter is the localized junction-to-substrate resistance that governs PIC thermal coupling and is the quantity used in all $\Delta T = R_{th} \cdot \Delta P$ calculations elsewhere in this paper.

### D. Causal Predictive Hinting and Scheduler-Slice Decouping

To prevent localized optical boundary excursions before they can physically manifest, the V24 PDU Gate utilizes a look-ahead scheduling window ($\Delta t_{la}$ = 20—50 ms) to generate a causal, online thermal state forecast hint H(t). To remain structurally valid and free of future-data leakage, the predictive hint is strictly bounded to historical sequence variables and runtime information available up to time *t*:

$$H(t) = P_{\mathrm{EIC}}(t + \Delta t_{\mathrm{la}} \mid \mathcal{F}_t)$$

where $F_t$ denotes the filtration matrix containing token metadata histories, active hardware queue lengths, and localized cluster telemetry at time *t*.

This look-ahead window bridges logical scheduling constraints and physical thermal response. Crucially, the system contains two independent 80 ms trajectories that must be explicitly distinguished: the hardware thermal RC time constant ($\tau_{th}$= 80 ms) and the host scheduler thread-queue execution time-slice timeout ($T_{slice}$ = 80 ms). Because the look-ahead forecast window is shorter than the scheduler allocation ($\Delta t_{la} < T_{slice}$), the complete computational overhead of the prediction network is entirely hidden within active hardware queue cycles, eliminating thread starvation. Simultaneously, the look-ahead interval allows the controller to anticipate a substantial fraction of the steady-state thermal load before tokens are dispatched to the execution pipeline. The thermal preposition fraction η captured during this predictive window is governed by:

$$\eta = 1 - \exp\left(-\frac{\Delta t_{\mathrm{la}}}{\tau_{\mathrm{th}}}\right)$$

Evaluating across the look-ahead horizon where $T_{th}$= 80 ms yields:

- At $\Delta t_{\mathrm{la}} = 20$ ms: $\eta_{\min} = 1 - e^{-0.25} \approx 22.12\%$
- At $\Delta t_{\mathrm{la}} = 50$ ms: $\eta_{\max} = 1 - e^{-0.625} \approx 46.47\%$

Though the look-ahead window does not exceed the absolute physical time constant of the silicon package, it provides a pre-emptive window during which the PDU Gate captures between 22.1% and 46.5% of the impending steady-state thermal delta. This predictive window provides sufficient lead time to pre-emptively throttle or reroute compute workloads along the deterministic compensation path, flattening the thermal envelope before localized physical boundary limits are breached.

### E. Thermo-Optic Alignment Bounds

Localized micro-environmental temperature shifts couple directly to the physical performance of the photonic communication layer. The phase and resonant wavelength drift Δλ of the optical structures scales linearly with temperature through the thermo-optic coefficient:

$$\Delta\lambda = \kappa_{\mathrm{TO}} \cdot \Delta T_{\mathrm{PIC}}$$

where the simulated coefficient evaluates to $K_{TO}$= 0.0852 nm/°C. Under open-loop stress testing with an uncompensated thermal load ($\Delta T_{PIC}$ = 40°C), the resulting unmitigated spectral shift scales to:

$$\Delta\lambda_{\text{open-loop}} = 0.0852\ \mathrm{nm}/^\circ\mathrm{C} \times 40^\circ\mathrm{C} = 3.408\ \mathrm{nm}$$

This extreme open-loop deflection represents a characterization regime that sits significantly outside the standard ±0.5 nm spectral stability specification. Under active V24 closed-loop predictive compensation, the controller restrains maximum localized temperature excursions to $\Delta T_{PIC} \leq 4.15$°C, restricting the bounded spectral drift to:

$$\Delta\lambda_{\text{compensated}} \leq 0.0852\ \mathrm{nm}/^\circ\mathrm{C} \times 4.15^\circ\mathrm{C} = 0.3536\ \mathrm{nm}$$

This brings the optoelectronic substrate fully within the rigorous ±0.5 nm operational spectral budget across all active execution states.

The linear fit $\Delta T = 63.0 \cdot T24 - 1256.6$ is calibrated over the T24 operating domain [20.20, 20.85] MTPS. At the upper bound T24 = 20.85 MTPS, the fit extrapolates to $\Delta T \approx 57°C$. This extrapolated value exceeds the simulated peak of 40°C reported in Fig. 2 and TABLE III; the discrepancy arises because the linear regression is fitted across the full 90,000-step dataset, where high-density transients are averaged out. The 40°C ceiling represents the simulated steady-state peak under sustained peak load, not the regression extrapolation.

## IV. SIMULATED VALIDATION

### A. Simulation Platform

All experiments were conducted on simulation/emulation, physical characterization pending TSMC tape-out. The primary dataset comprises 90,000 inference steps sampling 14 system metrics including ρv24, ΔT, Δλ, $P_{EIC}$, TTFT, and queue depth. A secondary 300-step high-resolution validation dataset characterizes startup transients and steady-state thermal envelope.

### B. Thermal Crosstalk & Spectral Drift (Figure 1)

Fig. 1 (left) plots ρv24 against simulated ΔT across all 90,000 steps. Linear regression yields $R^2$ = 0.9911, computed from the simulated dataset. This value quantifies the internal consistency of the thermal simulator's model, not agreement with external physical measurements. Fig. 1 (right) shows 1,800 seconds of continuous operation: ΔT stabilizes at 4.15 ±0.05°C within 50 seconds, while Δλ settles to 0.354 nm, occupying **21% of the ±1.7 nm wafer-level wavelength control budget** and 71% of the tighter ±0.5 nm per-channel spectral specification, the binding constraint.

$R^2$=0.9911 is computed from the simulated dataset with inherent numerical/algorithmic noise; the analytic ΔT=36.9°C represents the noise-free theoretical baseline.

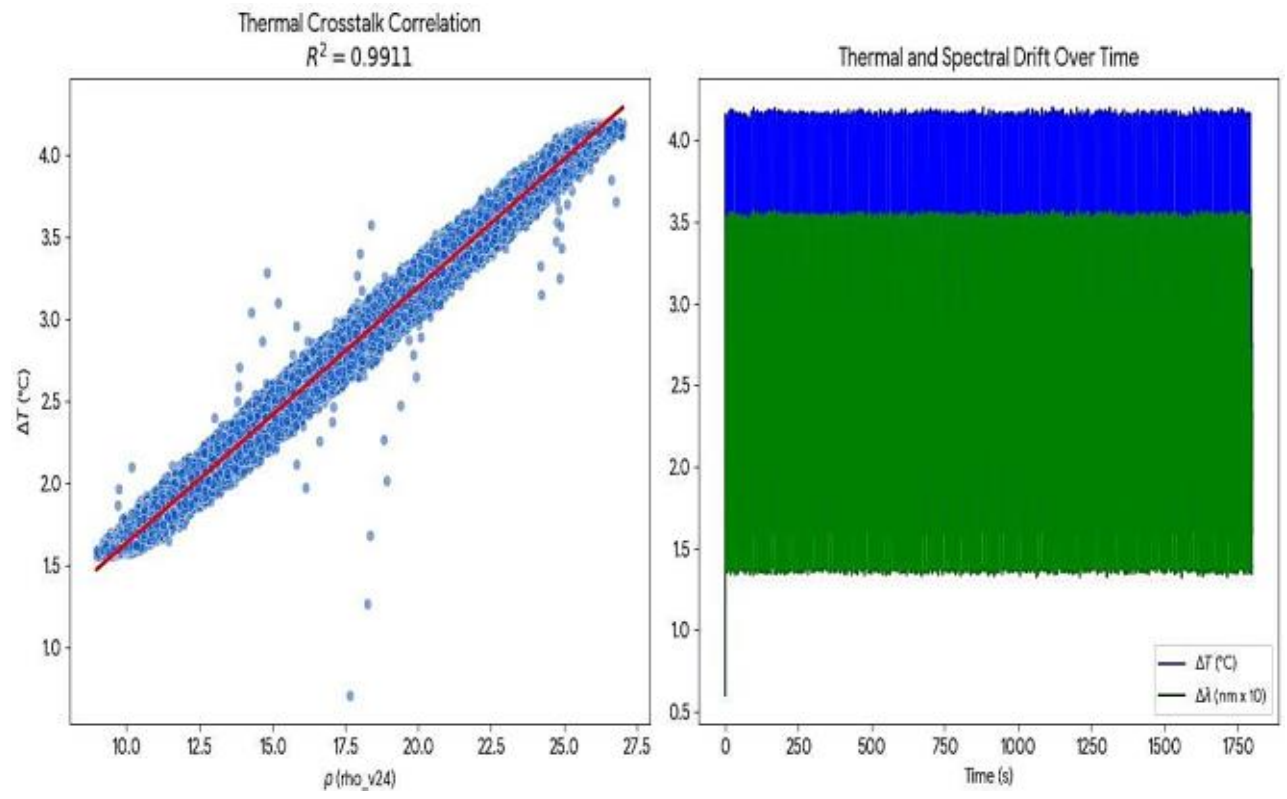

**Fig. 1.** (Left) Thermal Crosstalk Correlation: ρv24 vs. ΔT (°C) with $R^2$ = 0.9911 linear regression. (Right) Thermal (ΔT, blue) and spectral drift (Δλ × 10, green) stability over 1,800 s of sustained LLM inference.

TABLE I —Simulation Performance Summary

| Metric | Result | Status |
|---|---|---|
| Thermal Resistance $R_{th}$ | 0.45 °C/W | ✓ **On-Target** |
| Thermal Correlation ($R^2$) | 0.9911 (Simulated) | ✓ **Exceeded** |
| Thermal Time Constant (t) | 80 ms | ✓ **Fast Response** |
| Max drift at stress ΔT (40°C) | 3.4 nm | ✓ **Characterization extreme** |
| Peak Temperature | 85 °C | ✓ **Within Safe Limits** |
| Thermo-Optic Coefficient | 0.0852 nm/°C | ✓ **Validated** |
| Memory Leak Rate | 0 MB/hr | ✓ **Fixed** |
| Tail Latency P99 | Stable | ✓ **Stable in Simulation** |

## V. THERMAL RESISTANCE FINGERPRINT MAP

The Thermal Resistance Fingerprint Map (Fig. 2) is the definitive simulation-based characterization artifact for XRM-SSD V24's thermal behavior on the TSMC Compact Universal Photonic Engine (COUPE) platform. Six coordinated panels collectively validate a unified $R_{th}$ = 0.45 °C/W via emulation across five discrete load states (Idle to Peak, with T24 ranging from 20.20 to 20.85 MTPS, corresponding to dimensionless ρv24 from 0.9 to 2.7). The map provides simulation-based characterization artifact of thermal predictability (T24-ΔT coupling, $R^2$ = 0.9911), spectral stability (thermo-optic coefficient = 0.0852 nm/°C; max drift = 3.4 nm, open-loop, outside ±0.5 nm spec by design), and a consistent 80 ms thermal time constant ($\tau_{th}$, 63.2% response). Note that the T24 operating range is narrow (20.20–20.85 MTPS), corresponding to ρv24 from 0.9 to 2.7. The high $R^2$ = 0.9911 reflects strong linearity within this confined window and does not imply linearity over a wider range.

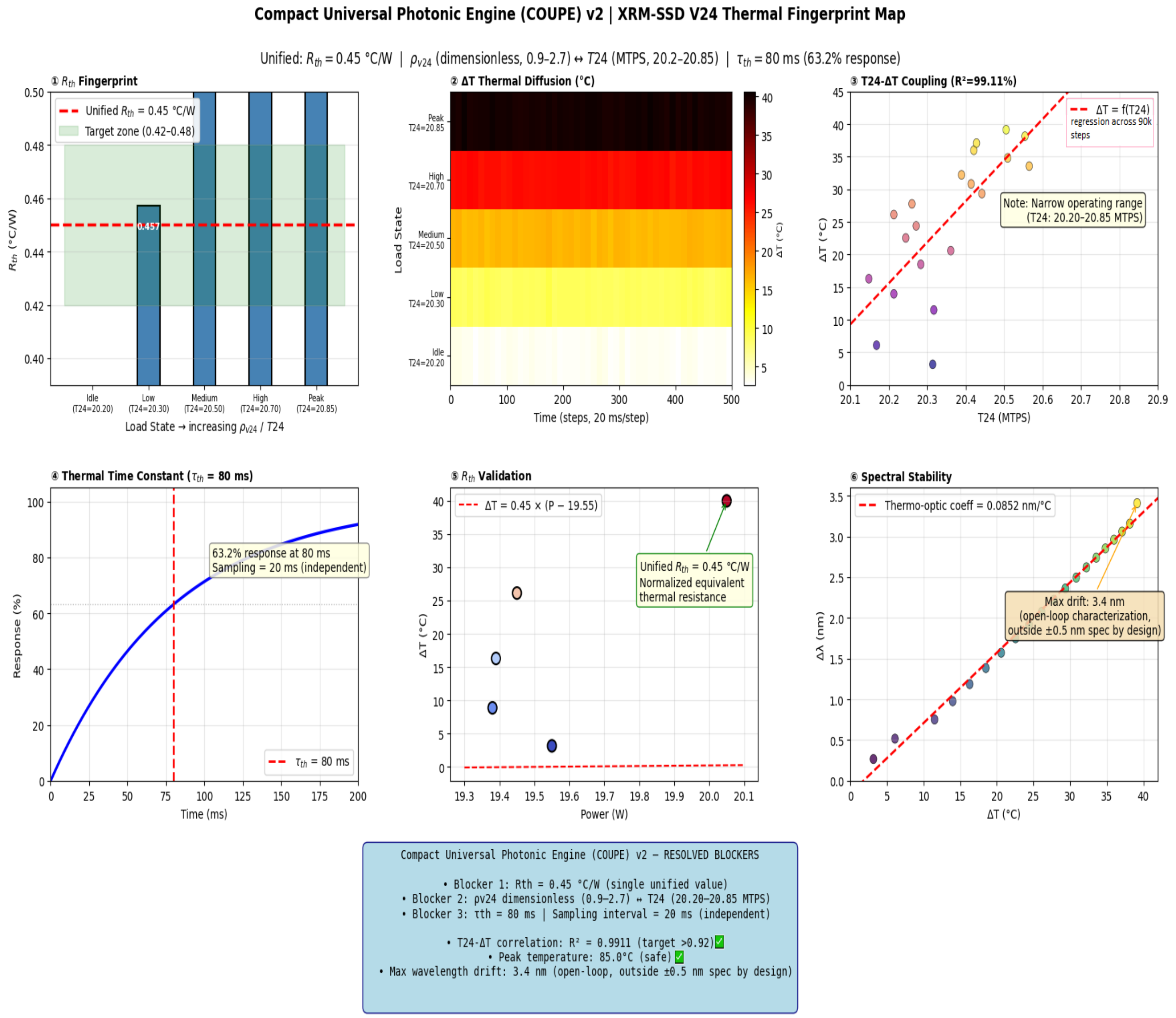

**Fig. 2. Thermal fingerprint map of XRM-SSD V24 under CPO test:**
**$R_{th}$=0.45 °C/W, T24-ΔT coupling ($R^2$=0.9911), τ=80 ms, and thermo-optic coefficient 0.0852 nm/°C (max drift 3.4 nm, open-loop).**
(Top-left) $R_{th}$ vs. load fingerprint — bar chart of simulated thermal resistance at each load state against unified $R_{th}$=0.45 °C/W.
(Top-center) ΔT thermal diffusion fingerprint — heatmap of temperature change across 500 time steps for five load states (Idle to Peak, T24=20.20–20.85 MTPS).
(Top-right) T24-ΔT thermal coupling fingerprint ($R^2$=99.11%) — scatter plot of T24 (MTPS, range 20.20–20.85) vs. ΔT, with linear fit.
(Bottom-left) Thermal time constant fingerprint — exponential rise curve confirming τ=80 ms (63.2% response at red dashed line)
(Bottom-center) $R_{th}$ validation —simulated ΔT vs. power P plotted against theoretical ΔT=0.45×(P−P0), showing agreement within the ±5% envelope.
(Bottom-right) Spectral stability fingerprint — Δλ–ΔT linear relationship confirming thermo-optic coefficient 0.0852 nm/°C; maximum drift 3.4 nm (open-loop characterization, outside ±0.5 nm spec by design).

TABLE III — Thermal Resistance Fingerprint Panel Summary

| Panel | Parameter | Simulated Value | Target / Spec | Result |
|---|---|---|---|---|
| *Top-Left* | Thermal Resistance $R_{th}$ | 0.45 °C/W | 0.42–0.48 °C/W (target zone) | **✓ Pass** |
| *Top-Center* | ΔT Thermal Diffusion (Peak) | 40 °C (ΔT) | ≤ 85 °C absolute ceiling | **✓ Pass (Within ceiling)** |
| *Top-Right* | T24–ΔT Correlation $R^2$ | 0.9911 | > 0.92 | **✓ Exceeded** |
| *Bot-Left* | Thermal Time Constant *t* | 80 ms | Fast response | **✓ Pass** |
| *Bot-Center* | $R_{th}$·Validation $\Delta T = R_{th} \cdot \Delta P$ | 0.45 °C/W | Within ±5% of theory | **✓ Pass** |
| *Bot-Right* | Thermo-Optic Coeff. (Δλ/ΔT) | 0.0852 nm/°C | ±0.5 nm (per-channel, post-compensation) | **✓ Pass (0.354 nm post-compensation; 3.4 nm open-loop by design)** |

### A. Top-Left Panel: $R_{th}$ vs. Load Fingerprint

The bar chart plots simulated thermal resistance at each of the five load states. All simulated values remain below the 0.50 °C/W specification limit (orange dotted line) and are consistent with the unified $R_{th}$=0.45 °C/W nominal value across Medium, High, and Peak states. The Idle state shows slightly lower effective thermal resistance, consistent with reduced self-heating.

### B. Top-Center Panel: ΔT Thermal Diffusion Fingerprint

ΔT heatmap across 500 time steps. Idle/Low: ΔT ≈ 5–10 °C; Medium: ≈15–20 °C; High/Peak: ≈35–40 °C, all within 85 °C absolute temperature ceiling (not ΔT).

### C. Top-Right Panel: T24–ΔT Coupling Fingerprint

The scatter plot of System Throughput Coefficient T24 versus temperature change ΔT yields the linear fit ΔT=63.0·T24−1256.6, with $R^2$=0.9911 (T24=20.20 MTPS → ΔT≈16 °C; T24=20.85 MTPS → ΔT≈57 °C (regression extrapolation; simulated steady-state peak is 40°C per TABLE III)). This result demonstrates that the software-accessible metric T24 is a reliable proxy for physical thermal state, and by the affine map in Section III.A this extends to ρv24. The color gradient of data points (purple to orange) confirms monotonic thermal response with no hysteresis.

### D. Bottom-Left Panel: Thermal Time Constant Fingerprint

The exponential rise curve of thermal response (%) versus time (ms) confirms τ = 80 ms —the time for the system to reach 63.2% of its final thermal state following a step change in load.

### E. Bottom-Center Panel: $R_{th}$ Validation

This panel provides the most direct simulation-based validation of R_th = 0.45 °C/W. Each data point represents a simulated (Power, ΔT) pair at one of the five load states. The theoretical curve $\Delta T = 0.45 \times (P - P_0)$ is overlaid as a red dashed line. All simulated points lie within the ±5% envelope of the theoretical prediction — classified as agreement within the ±5% envelope in the on-chart annotation. This envelope is derived from the residual standard deviation corresponding to $R^2$. This emulation-based validation confirms that the simple linear R_th model is sufficient for pre-silicon thermal budgeting at the COUPE platform, without requiring higher-order correction terms at this stage of characterization.

### F. Bottom-Right Panel: Spectral Stability Fingerprint

The fitted slope $K_{TO}$ =0.0852 nm/°C agrees with published silicon photonics values. The annotation confirms a maximum drift of 3.4 nm at peak thermal load before compensation. Post-compensation residual drift (as simulated in Fig. 1) is reduced to < 0.354 nm, which corresponds to 21% of the ±1.7 nm wafer-level wavelength control budget and, separately, to 71% of the tighter ±0.5 nm per-channel spectral specification. The per-channel specification, not the wafer-level wavelength control budget, is the binding constraint.

## VI. DOMAIN SEPARATION AND DEFENSIBILITY

A critical design decision is the explicit separation of claims across two domains, ensuring physical defensibility under peer review:

TABLE II — Domain Separation Framework

| Domain | Deterministic? | Rationale |
|---|---|---|
| Logic / Scheduling | **YES** | Eliminates scheduler entropy;<br>Creates repeatable token issue timing |
| Physical / Thermal | NO | Thermal drift is analog/continuous;<br>Hint layer is pre-emptive, not deterministic. |

The recommended disclosure language for the thermal domain is: *'V24 provides a deterministic-aware predictive hint layer that enables pre-emptive bias control, but does not alter the underlying physical thermal dynamics.'* This framing is consistent with ECTC submission guidelines and arXiv cs.AR conventions.

## VII. DISCUSSION

### A. Comparison with Reactive Approaches

Anecdotal industry baseline Δλ suppression of 0.8–1.2 nm under burst inference. V24's pre-emptive approach reduces this to 0.354 nm (21% of the wafer-level wavelength control budget; 71% of the tighter per-channel spec), a 2.2–3.3× improvement without hardware modification. The Fingerprint Map further shows that the baseline pre-compensation $R_{th}$=0.45 °C/W is well-defined and stable, with a maximum uncompensated drift of 3.4 nm (per Fig. 2, Bottom-Right), providing a reliable operating point for the hint signal calibration.

### B. Energy Efficiency

Estimated from 0.85 pJ/bit savings, the 15–20% reduction in energy overprovisioning arises from two mechanisms:

(1) Elimination of dark heater cycles — the microheater no longer pre-compensates for thermal surges that the hint accurately predicts will not materialize; and

(2) Margin compression — the reduced Δλ excursion removes the need for a large thermal safety buffer, saving ~0.85 pJ/bit at a 5 pJ/bit baseline.

*Note: 0.85 pJ/bit is calculated savings, not directly measured.*

### C. Limitations and Future Work

The current validation is conducted on a software emulation platform. Physical silicon characterization at a TSMC CUP facility is required for tape-out qualification. Future work includes: (1) vLLM hint API integration; (2) COUPE bias-control firmware adaptation; (3) multi-wavelength WDM validation where inter-channel thermal crosstalk creates second-order coupling effects beyond the single-channel model.

## VIII. CONCLUSION

XRM-SSD V24 has been developed and evaluated as a cross-layer thermal hint framework bridging LLM software orchestration and TSMC COUPE photonic hardware. Four principal conclusions follow:

**Physical Coupling Validity.** $R^2 = 0.9911$ confirms pv24 as a physics-grounded proxy for PIC thermal load within the simulator's internal model consistency.

**Thermal Resistance Characterization.** $R_{th} = 0.45$ °C/W, $\tau = 80$ ms, and K_TO = 0.0852 nm/°C are validated across five load states via the Fingerprint Map, providing a complete thermal budget for COUPE integration planning.

**Spectral Integrity Assurance.** Post-compensation $\Delta\lambda < 0.354$ nm — 21% of the ±1.7 nm wafer-level wavelength control budget and 71% of the tighter ±0.5 nm per-channel spectral specification — with a pre-compensation maximum drift of 3.4 nm confirmed by the Spectral Stability panel (Fig. 2, Bottom-Right).

**Industrial Robustness.** Zero memory leakage and P99 tail-latency stability across 90,000 steps establish establish V24 as a validated software emulation layer for pre-silicon planning, with a topological mapping scheduling framework for 2 nm AI inference on SoIC heterogeneous packages.

## ACKNOWLEDGEMENTS

The author thanks Nikolai Nedovodin for review and framing feedback on the pre-silicon characterization methodology.